\documentclass[12pt]{amsart}

\usepackage{amssymb}
\usepackage{epsfig}
\usepackage{comment}
\usepackage{amsmath}
\usepackage{caption}
\usepackage[section]{placeins}
\usepackage{graphicx}
\usepackage{units}
\usepackage{color}
\usepackage{subfig}
\usepackage{threeparttable}
\usepackage{bm}
\usepackage[margin=1.0in]{geometry}

\graphicspath{ {Figures/} }

\theoremstyle{definition}

\numberwithin{equation}{section}
\theoremstyle{remark}

\begin{document}

%\title[ ]{Molecular dynamics study of hydrogen-defect interactions in palladium nanoparticles}
%\title[ ]{Atomistic simulation of solute-defect interactions in palladium hydrides across multiple time scales}
\title[]{Exploring Solute-Defect Interactions in Nanosized Palladium Hydrides across Multiple Time Scales}

\author[]
{Xingsheng~Sun${}^\dagger$}

\address
{
    Department of Mechanical and Aerospace Engineering,
    University of Kentucky,
    Lexington, KY 40506, USA.
    ${}^\dagger$ Corresponding author. E-mail address: xingsheng.sun@uky.edu.
}

\begin{abstract}
We employ two different atomistic methods to investigate solute-defect interactions in nanosized palladium-hydrogen (Pd-H) systems across multiple time scales. The first method, referred to as Diffusive Molecular Dynamics (DMD), focuses on capturing hydride phase transformation and the evolution of solute-induced lattice defects over a diffusive time scale. The second method, Molecular Dynamics (MD), provides more detailed insights into atomic movements and lattice relaxation over the time scale of thermal vibrations. These two methods are connected with MD simulations initialized using statistical measures of microscopic variables obtained from DMD at different H/Pd ratios. Our study demonstrates that DMD effectively captures the propagation of an atomistically sharp hydride phase boundary as well as the dynamics of solute-induced misfit dislocations and stacking faults. While the  H-concentrated phase leads to a reduction in the vibrational energy, the presence of stacking faults locally increases the vibrational energy of both Pd and H atoms. Furthermore, the MD simulation results align with DMD in terms of equilibrium potential energy, the preservation of hydride phase boundary, and the spatial distribution of stacking faults. We thoroughly characterize the lattice crystalline structures in four key regions of the particle. We observe a preference for H atoms to occupy tetrahedral interstitial sites near stacking faults due to the lower stacking fault energies provided by these sites within the H-concentrated phase.

\end{abstract}

\maketitle


\section{Introduction}
\label{sec:intro}

Diffusion of interstitial solutes in crystalline solids is of broad technological relevance and importance in various energy applications including lithium-ion batteries~\cite{mendez2018diffusive} and hydrogen storage systems~\cite{sun2019atomistic}. During this process, solute atoms migrate from one interstitial site to an adjacent vacant site within the host lattice,  contributing to energy storage, release, and conversion. Concurrently, the host crystal undergoes lattice deformation or phase transformation to accommodate the solute atoms. These changes in lattice structure can result in the formation of lattice defects, including dislocations and stacking faults, which profoundly influence both the transport properties of solutes and the mechanical properties of the host materials. Consequently, comprehending the nanomechanisms underlying solute-defect interactions is crucial for material designers. By understanding how solutes modify crystal structures and how defects impact solute diffusion at the nanoscale, valuable insights can be gained to enhance material properties (e.g., mechanical, chemical, electrical) and mitigate material failures.

Palladium hydride ($\text{PdH}_x$) has served as a prototypical system for the study of solute-induced phase transformations and solute-defect interactions in both fundamental and applied research fields for several decades~\cite{okuyama1998path, li2014hydrogen, zalineeva2017octahedral, narayan2017direct, ulvestad2015avalanching, ulvestad2017three, sytwu2018visualizing}. Pd and its nanomaterials are particularly noteworthy due to their exceptional ability to absorb a significant amount of hydrogen at readily achievable temperatures and pressures, with hydrogen atoms exhibiting high mobility within the Pd lattice. The phase diagram of $\text{PdH}_x$ reveals a dilute $\alpha$ phase ($x<0.015$ at room temperature) and a concentrated $\beta$ phase ($x>0.6$ at room temperature). While the face-centered cubic (FCC) structure is maintained by the Pd lattice in both phases, the $\alpha$ to $\beta$ phase transformation is accompanied by a $4\%$ increase in the lattice constant, potentially leading to the formation of lattice defects such as misfit dislocations and stacking faults.

The dynamics of H atoms in nanostructured Pd have been extensively studied using experimental methods. For instance, under gas-phase conditions, the hydrogenation of Pd nanoparticles involves three steps~\cite{li2014hydrogen, zalineeva2017octahedral, sun2019atomistic}: (1) the dissociation of H\textsubscript{2} into H atoms on the surface of the Pd nanoparticles (i.e., adsorption), (2) the saturation of H atoms into a subsurface layer, and (3) the diffusion of H atoms into the interior, occupying the octahedral interstitial sites of the FCC lattice and forming a H-rich $\beta$ phase (i.e., absorption). A few recent experiments have suggested that in individual nanosized particles, the absorption is characterized by the propagation of a hydride phase boundary with a thickness of several atomic layers~\cite{narayan2017direct, ulvestad2015avalanching, ulvestad2017three, sytwu2018visualizing}. Due to the rate-limiting nature of subsurface saturation~\cite{okuyama1998path, wilde2008penetration}, the speed of phase boundary movement can be as low as $1~\text{nm/s}$, resulting in long time scales for the completion of the hydrogenation process. In addition to the overall transformation mechanism and the long-term transformation process, other intriguing examples of solute-defect interactions in Pd-H systems include dislocation healing by phase transformation~\cite{ulvestad2017self}, slow-down of phase transformation due to grain growth~\cite{alekseeva2021grain}, shear strain inversion during phase transformation~\cite{ulvestad2015avalanching}, enhanced H migration via dislocation pipe diffusion~\cite{heuser2014direct}, and phase diagram hysteresis caused by defect-interface interactions~\cite{weadock2021interface}.

The aforementioned experimental studies highlight the dependence of H diffusion in Pd nanosized particles on physical processes occurring across a broad range of time scales. These processes span from the femtosecond timescale of thermal vibrations of Pd and H atoms to the propagation of hydride phase boundaries in Pd nanoparticles, which can take seconds or even longer. The convergence of atomistic length scales and multiple time scales presents a significant challenge for modeling and simulation. Density Functional Theory (DFT) calculations have been widely employed to predict formation energy, activation barrier, and vibrational energy within crystal surfaces~\cite{paul1996density, dong1998hydrogen, johansson2010hydrogen} as well as dislocation cores~\cite{lawler2010first, schiavone2016ab}. Meanwhile, the thermal vibrations of H atoms in adsorption and absorption dynamics are typically studied through Ab Initio~\cite{gross2007hydrogen, gross2010ab} or classical~\cite{lozano2010molecular, zhou2018temperature} Molecular Dynamics (MD) simulations. To overcome the time scale limitation of MD simulations, a variety of techniques have been developed, including transition state theory based accelerated MD methods~\cite{voter1998parallel, voter1997method, so2000temperature}, kinetic Monte Carlo (MC) methods~\cite{henkelman2001long}, phase field crystal methods~\cite{berry2008melting}, atomic density function methods~\cite{jin2006atomic}, and quasi-particle approaches~\cite{demange2022atomistic}. The applicability of these time-acceleration methods to model solute-defect interactions over long diffusive time scales, however, has not been demonstrated.

In this work, we employ a recently developed computational method, referred to as Diffusive Molecular Dynamics (DMD), to investigate the intricate dynamics of hydride phase transformation in Pd nanoparticles. DMD represents a novel approach for simulating long-term diffusive mass transport and heat transfer while maintaining atomic-level resolution~\cite{kulkarni2008variational, venturini2014atomistic, li2011diffusive, sun2017acceleration}. To bridge the significant timescale gap, the underlying assumption of DMD theory is that on an intermediate timescale, much larger than that of thermal vibrations, the microscopic state variables of an atomic site---such as position, momentum, and occupancy---can be treated as random variables~\cite{sun2019atomistic}. Then the microscopic state of the system is characterized by some probability density functions, and this intermediate timescale aligns with one time-step of DMD simulation. Consequently, the DMD theory aims to solve the statistics of these random variables over a long time scale. Specifically, DMD combines a discrete kinetic model governing the evolution of local solute concentrations at sites over diffusive timescales with a nonequilibrium statistical thermodynamics model that relaxes the crystal structure and provides the necessary driving forces for kinetics. DMD has already demonstrated its applicability in various areas, such as H diffusion in Pd nanoparticles~\cite{sun2018long, sun2019atomistic} and nanofilms~\cite{sun2016deformation, wang2015long}, nanoindentation and sintering of copper~\cite{li2011diffusive}, nanovoid growth in copper~\cite{ariza2012hotqc, ponga2015finite} and aluminum~\cite{ponga2017voids}, solute-defects interaction in aluminum-magnesium~\cite{dontsova2015solute, dontsova2014solute}, heat conduction in silicon nanowires~\cite{venturini2014atomistic}, lithiation of silicon nanopillars~\cite{mendez2018diffusive}, among others. 

In this paper, we synergistically combine Diffusive Molecular Dynamics (DMD) with classical Molecular Dynamics (MD) to investigate interactions between H solutes and lattice defects in Pd nanoparticles, achieving atomic resolution across multiple time scales. The broad time window of DMD allows us to effectively study hydride phase transformation and the evolution of solute-induced lattice defects over diffusive time scales. On the other hand, classical MD provides more detailed insights into atomic movements and lattice relaxation on the timescale of thermal vibrations. These two complementary methods are interconnected by initializing MD simulations using the statistical measures of microscopic variables obtained from DMD at different H/Pd ratios. 

The remainder of this paper is structured as follows. Section~\ref{sec:method} presents a concise overview of the theory, model equations, and setup for both DMD and MD simulations. Subsequently, in Section~\ref{sec:result}, we discuss the results of our numerical experiments, encompassing the motion of the hydride phase boundary, the dynamics of H-induced stacking faults, the impact of hydride phases and stacking faults on vibrational energy, the comparisons between MD and DMD simulations, the influence of hydride phases on lattice crystallinity, and the effect of defects on H-preferred interstitial sites. To explain the observed energetically stable sites near stacking faults, we employ Molecular Statics to calculate the intrinsic stacking fault energy arising from different hydride phases and interstitial sites. Finally, in Section~\ref{sec:concl}, we provide a summary of our findings and present concluding remarks.

\section{Methodology}
\label{sec:method}

For the sake of completeness, in this section we first briefly summarize the DMD theory and its implementation for simulating H diffusion in Pd nanoparticles. Further details and applications of DMD can be found in Ref.~\cite{kulkarni2008variational, li2011diffusive, venturini2014atomistic, sun2017acceleration, simpson2016theoretical}. Then we introduce the setup of MD simulations that will be used to explicitly solve the displacive atomic movements.

\subsection{Diffusive Molecular Dynamics}

We focus our attention on a three-dimensional Pd-H system which comprises two types of sites, i.e., host and interstitial sites. Each host lattice site is always occupied by a Pd atom, whereas each interstitial site can be either occupied by an H atom or vacant. We denote the sets of host and interstitial sites by $I_\text{Pd}$ and $I_\text{H}$, respectively. The number of host and interstitial sites are denoted by $N_\text{Pd}$ and $N_\text{H}$, respectively, i.e., $N_\text{Pd}=|I_\text{Pd}|$ and $N_\text{H}=|I_\text{H}|$. At each interstitial site $i\in I_\text{H}$, we define an occupancy function $n_i$ as~\cite{sun2019atomistic}
\begin{equation}
n_i =
 \begin{cases}
    1      & \text{if the site } i \text{ is occupied by a H atom},\\
    0      & \text{if the site } i \text{ is vacant}.\\
  \end{cases}
  \label{eq:occ}
\end{equation}
By contrast, the occupancy of the host sites is always $1$, i.e.,  $n_i =1,~i\in I_\text{Pd}$. It follows from the definition Eq.~(\ref{eq:occ}) that the occupancy array $\{n\}$ takes values in a set consisting of the elements of $\{0,1\}$. We refer to this set as $\mathcal{O}$, defined by
\begin{equation}
\mathcal{O} = \{0,1\}^{N_\text{H}}
\end{equation}

In addition to occupancy, we denote the instantaneous position and momentum of site $i$ by $\bm{q}_i$ and $\bm{p}_i$, respectively. When viewed on time scales much longer than that of atomic vibrations, these microscopic state variables can be regarded as random variables that have a joint probability distribution characterized by density function $\rho\big(\{\bm{q}\}, \{\bm{p}\}, \{n\}\big)$, where $\{\bm{q}\}=\{\bm{q}_{i}:i\in I_\text{Pd}\cup I_\text{H}\}$, $\{\bm{p}\}=\{\bm{p}_{i}:i\in I_\text{Pd}\cup I_\text{H}\}$ and $\{n\}=\{n_{i}:i\in I_\text{H}\}$~\cite{sun2019atomistic}. Then we can define the expectation or macroscopic value of any quantity $A\big(\{\bm{q}\}, \{\bm{p}\}, \{n\}\big)$ via the widely used phase average in the classical statistical mechanics~\cite{kulkarni2008variational}
\begin{equation} 
\langle A \rangle
= \sum_{\substack{\{n\}\in\mathcal{O}}}
\dfrac{1}{h^{3(N_\text{Pd}+N_\text{H})}}
\int A\big(\{\bm{q}\}, \{\bm{p}\}, \{n\}\big) \rho\big(\{\bm{q}\}, \{\bm{p}\}, \{n\}\big)
\prod_{i\in{I_\text{Pd}} \cup I_\text{H}}\text{d}\bm{q}_i\text{d}\bm{p}_i,
\label{eq:ave}
\end{equation}
where $h$ is the Planck's constant. Following the Gaussian density clouds of atomic positions~\cite{li2011diffusive, dontsova2014solute, dontsova2015solute}, we further assume that $\bm{q}_i$ and $\bm{p}_i$ of each host and interstitial sites are characterized by the normal distribution, whereas $n_i$ of each interstitial site follows Bernoulli distribution. Then the probability density function $\rho$ can be shown as~\cite{sun2017acceleration}
\begin{equation} 
\begin{aligned}
\rho \big( \{\bm{q}\}, \{\bm{p}\}, \{n\} \big)  =
& \dfrac{1}{\mathnormal{\Xi}} \exp\Bigg(-\sum_{i\in{I_\text{Pd} \cup I_\text{H}}}
\bigg(\frac{1}{2\sigma_i^2}|\bm{q}_i - \bar{\bm{q}}_i|^2
+ \frac{1}{2k_{\text{B}}T_i m_i}|\bm{p}_i - \bar{\bm{p}}_i|^2 \bigg) \\
& + \sum_{i\in{I_\text{H}}}
n_i\log{\frac{x_i}{1-x_i}} \Bigg),
\end{aligned}
\label{eq:trialP_MH}
\end{equation}
with the partition function
\begin{equation} 
\mathnormal{\Xi} =
\prod_{i\in{I_\text{Pd}} \cup I_\text{H}}
\bigg(\dfrac{\sigma_i\sqrt{k_{\text{B}}T_i m_i}}{\hbar}\bigg)^3
\prod_{i\in{I_\text{H}}}
\dfrac{1}{1-x_i},
\label{eq:trialpartfun_MH}
\end{equation}
where $k_\text{B}$ is the Boltzmann constant and $\hbar$ is the reduced Planck's constant. $m_i$ and $T_i$ are the atomic mass and absolute local temperature, respectively. $\bar{\bm{q}}_i$, $\sigma_i$, $\bar{\bm{p}}_i$, and $x_i$ are parameters characterizing the probability density function. In comparison with equilibrium statistical thermodynamics, we can also regard $\rho$ and $\Xi$ as the nonequilibrium grand-canonical probability distribution and partition function, respectively. Moreover, one can derive the following equations using Eqs.~(\ref{eq:ave}) and (\ref{eq:trialP_MH})
\begin{equation}
\bar{\bm{q}}_i = \langle \bm{q}_i \rangle, \quad i\in I_\text{Pd}\cup I_\text{H},
\end{equation}
\begin{equation}
3\sigma^2_i = \langle \bm{q}^2_i \rangle - \langle \bm{q}_i \rangle^2, \quad i\in I_\text{Pd}\cup I_\text{H},
\end{equation}
\begin{equation}
\bar{\bm{p}}_i = \langle \bm{p}_i \rangle, \quad i\in I_\text{Pd}\cup I_\text{H},
\end{equation}
\begin{equation}
3k_{\text{B}}T_i m_i = \langle \bm{p}^2_i \rangle - \langle \bm{p}_i \rangle^2, \quad i\in I_\text{Pd}\cup I_\text{H},
\end{equation}
and
\begin{equation}
x_i = \langle n_i \rangle, \quad i\in I_\text{H}.
\end{equation}
In other words, $\bar{\bm{q}}_i$ and $\sigma_i$ are the mean and standard deviation (SD) of $\bm{q}_i$, respectively. $\bar{\bm{p}}_i$ and $\sqrt{k_{\text{B}}T_i m_i }$ are the mean and SD of $\bm{p}_i$, respectively. $x_i$, referred to as atomic faction, is the mean of $n_i$. One can also estimate the average vibrational frequency $\omega_i$ at site $i$ using 
\begin{equation}
\omega_i = \sqrt{\dfrac{k_\text{B}T_i}{m_i\sigma_i^2}}.
\label{eq:frequency}
\end{equation}
 
The statistics of microscopic variables can be determined using meanfield approximation for the free energy of the system. After straightforward deviations according to Ref.~\cite{sun2017acceleration}, we have $\bar{\bm{p}}_i=0$, and $\bar{\bm{q}}_i$ and $\sigma_i$ can be solved by minimizing the variational Gaussian Helmholtz energy
\begin{equation} 
\begin{aligned}
\min\limits_{ \{\bar{\bm{q}}\},\{\sigma\}}\mathcal{F} = 
& \langle{V}\rangle
+ \dfrac{3}{2}\sum_{\substack{i\in{I_\text{Pd}}}} k_{\text{B}}T_i\bigg(\log{\dfrac{\hbar^2}{k_{\text{B}} T_i m_i \sigma_i^2}} -1\bigg) \\
& + \dfrac{3}{2}\sum_{\substack{i\in{I_\text{H}}}} k_{\text{B}}T_i\bigg(\log{\dfrac{\hbar^2}{k_{\text{B}}T_i m_i \sigma_i^2}} + x_i - 2\bigg) \\
& + \sum_{\substack{i\in{I_\text{H}}}} k_{\text{B}}T_i\big(x_i\log{x_i}+(1-x_i)\log{(1-x_i)}\big),
\end{aligned}
\label{eq:fren0}
\end{equation}
where $V\big(\{\bm{q}\}, \{n\} \big)$ represents the interatomic potential energy. $\langle V \rangle$ can also be regarded as the {\it thermalization} of the interatomic potential, since it is computed by taking into account atomic vibrations and positions at certain temperatures. Moreover, it is noteworthy that based on the Gibbs-Bogoliubov inequality, the Helmholtz free energy estimated by Eq.~(\ref{eq:fren0}) provides an upper bound on the true Helmholtz free energy~\cite{venturini2014atomistic}.

The nonequilibrium thermodynamics model formulated in Eq.~(\ref{eq:fren0}) requires kinetic laws to describe the temporal evolution of the atomic fields, i.e., atomic temperature $T_i$ and atomic fraction $x_i$. In this work, we assume that the heat transfer is much faster than the mass transport. As a result, the temperature $T_i$, becomes uniform over all the sites, and is equal to a constant value $T$. Then we can formulate the exchange chemical potential at one interstitial site $i$ by differentiating Eq.~(\ref{eq:fren0}) with respect to $x_i$, i.e.,
\begin{equation}
\mu_i = \dfrac{\partial\mathcal{F}}{\partial x_i} =
\dfrac{3}{2} k_\text{B}T + k_\text{B}T\log{\dfrac{x_i}{1-x_i}}+\dfrac{\partial\langle{V}\rangle}{\partial{x_i}},\quad i \in I_\text{H}.
\label{eq:chempot}
\end{equation}

In this work, a discrete kinetic law is employed to govern slow mass transport at the atomistic length scale. At any time step, it enforces the balance of mass at each interstitial site, i.e.,
\begin{equation}
\dot{x}_i = \sum\limits_{j\neq i} J_{ij}, \quad i,j\in{I_\mathrm{H}},
\end{equation}
where $J_{ij}$ denotes the {\it bondwise} mass flux between site $i$ and site $j$. It is {\it antisymmetric} between the two sites and satisfies $J_{ij} = -J_{ji}$. Following Refs.~\cite{sun2019atomistic, sun2018long, sun2017acceleration, venturini2014atomistic}, we assume that the mass exchange $J_{ij}$ between two sites is governed by an {\it empirical} linear kinetic law
\begin{equation}
J_{ij} = 
\dfrac{B_0}{T} \dfrac{x_i + x_j}{2} (\mu_j-\mu_i), \quad i,j\in{I_\text{H}}, 
\label{eq:linear}
\end{equation}
where $B_0$ denotes the bondwise diffusion coefficient, and it can be calibrated to fit some experimental measures such as the speed of diffusion. Eq.~({\ref{eq:linear}}) shows that the mass flux between two sites is essentially driven by the gradient of the local chemical potential. If the difference of chemical potentials at two sites is zero, then these two sites will be in local equilibrium and the net mass exchange between them is zero. This is also consistent with the grand-canonical equilibrium thermodynamics. Moreover, this kinetic law was developed based on Onsager's theory of kinetic relations~\cite{de2013non, venturini2014atomistic}. It has been validated against the analytical solution of a three-dimensional random walk problem~\cite{sun2017acceleration}. Other applications include H diffusion in metallic nanoparticles~\cite{sun2019atomistic, sun2018long, sun2017acceleration} and nanofilms~\cite{venturini2014atomistic, wang2015long, sun2016deformation}, surface segregation in binary alloys~\cite{gonzalez2016numerical}, and spin diffusion in one-dimensional alloys~\cite{farmer2017spin}. 

A few details are highlighted regarding the implementation of the DMD method in this work. Firstly, the DMD method couples two problems, an optimization problem formulated in Eq.~(\ref{eq:fren0}), and a first-order ordinary differential equation (ODE) formulated in Eq.~(\ref{eq:linear}). We discretize the ODE using an explicit time integrator~\cite{sun2017acceleration}, and then at each time step solve the optimization problem using a quasi-Newton Broyden-Fletcher-Goldfarb-Shanno (BFGS) method~\cite{BFGS}. The initial guess of the optimization problem at one time step is set to the latest solution of the optimization solver to accelerate its convergence. See Ref.~\cite{sun2017acceleration} for details of the computation procedure. Secondly, the thermalization of interatomic potential $\langle V \rangle$ in Eq.~(\ref{eq:fren0}) can be solved by a few numerical methods such as Gaussian quadratures~\cite{kulkarni2008variational, venturini2014atomistic}, Graph Neural Networks~\cite{saxena2023gnn}, Gaussian function fitting~\cite{lesar1991thermodynamics, li2011diffusive}, and Monte-Carlo sampling~\cite{sun2017acceleration}. In this work, this average is calculated by Jensen's inequality with respect to $\{n\}$ and third-order Gaussian quadratures on a sparse grid with respect to $\{\bm{q}\}$~\cite{sun2017acceleration}, which provides an excellent compromise between computational cost and numerical error. Thirdly, the linear kinetic equation Eq.~(\ref{eq:linear}) is an {\it empirical} one. For the sake of simplicity, it is assumed that the model parameter $B_0$ does not depend on a site's local atomic environment. However, the local chemical potential $\mu_i$ in Eq.~(\ref{eq:chempot}) at each site does account for detailed atomic environments (i.e., structural and chemical configurations). Therefore, H diffusion is locally modified, e.g., near distorted lattice and a sharp phase interface. Finally, the summation in Eq.~(\ref{eq:linear}) is conducted over all diffusing neighbors. For simplicity, but without loss of generality, the DMD method only takes into account mass exchange within first nearest neighbors.

We simulate the absorption of H by a spherical Pd nanoparticle at the room temperature (i.e., $T=300~\text{K}$). The nanosphere, with a diameter of $20$~nm, consists of $261,563$ host sites and $261,742$ octahedral interstitial sites. The host lattice sites are fully occupied by Pd atoms, so the atomic fraction is always $1$ throughout simulations. The interstitial sites are either occupied by H atoms or vacant, so the atomic faction varies within the range $(0,1)$. Following previous studies~\cite{sun2017acceleration, sun2018long, sun2019atomistic}, we assume that the interstitial sites located within an outermost layer of the particle have already been in equilibrium with the surrounding H environment. The thickness of this layer is set to $0.5$ nm, based on the value provided by an equilibrium Monte Carlo method~\cite{ruda2010atomistic}. We also fix $x_i=1$ at all interstitial sites within this layer. The H fractions of other interstitial sites are initialized with $10^{-3}$, corresponding to the dilute $\alpha$ hydride phase. The lattice parameter of the particle is set to $a_\text{L}=3.885~\text{\AA}$ for the initial relaxation of the system. Then we employ the DMD model to predict the inward diffusion of H atoms from the outmost layer. An embedded atom method (EAM) potential~\cite{zhou2008embedded} is employed to capture the interactions between atoms. The diffusive parameter of the kinetic law is set to $B_0 = 500.0~\text{K/(eV}\cdot\text{s)}$, in line with previous studies~\cite{sun2019atomistic, sun2018long, sun2017acceleration}. The time step size for integrating the discrete kinetic equation is $2.0\times10^{-3}$~s. The total simulation time of H diffusion is $80.0$~s, which is sufficient for the complete phase transformation in the nanoparticle.

\subsection{Molecular Dynamics}
\label{sec:MD}

Large-scale molecular dynamics (MD) simulations are initialized by the DMD simulation results that aim to provide an energetically stable atomic configuration of both Pd and H atoms. Specifically, we use the DMD results at multiple time points with a broad range of H/Pd ratios. The DMD method states that, over a sufficiently long time scale, the instantaneous position and the occupancy of each site can be regarded as random variables following the normal and Bernoulli distributions, respectively. As a result, we use the normal distributions with the mean and SD calculated by the DMD simulation to randomly generate the initial positions of Pd atoms for the MD simulation. We further adopt the Bernoulli distributions with the mean also provided by DMD to randomly generate the initial occupancies of interstitial sites. If an interstitial site is occupied by an H atom, the initial position is then randomly generated based on the corresponding normal distribution. 

The MD simulations are performed using the Large-scale Atomic/Molecular Massively Parallel Simulator (LAMMPS) solver~\cite{plimpton1995fast}. We use the same EAM potential~\cite{zhou2008embedded} as employed in the DMD simulation. The temperature is fixed at $300$~K, using a Nos{\'e}-Hoover thermostat. The time-step size for solving Newton's equations of motion is $0.1$~fs, and the total simulation time for all simulations is $1.0$~ns. The total numbers of both Pd and H atoms are kept constant throughout the MD simulations. We find that the H-concentrated phase can lead to large variances of atomic positions. As a result, we employ the time-averaged atomic positions of Pd and H atoms over $0.01$~ns to identify the types of lattice structures and interstitial sites. In addition, the simulation results are visualized using OVITO~\cite{stukowski2010visualization}. The lattice structural types are identified by Common Neighbor Analysis~\cite{stukowski2012structure}, and the dislocation types are identified by Dislocation Extraction Algorithm~\cite{stukowski2012automated}.

\section{Results and discussions}
\label{sec:result}

\subsection{Evolution of statistics of microscopic variables} 

We initiate our investigation by analyzing H diffusion during the absorption process. To achieve this, we extract statistical information on microscopic variables at eight distinct steps from the DMD simulation. The corresponding H/Pd ratios and time points for each step are provided in Table~\ref{tab:atomnum}.

\begin{table}[!ht]
\centering
\begin{tabular}{ccccc}
\hline
\hline
 & \multicolumn{2}{c}{DMD simulation} & \multicolumn{2}{c}{MD simulation}    \\
 & Time (s) & H/Pd & Number of Pd atoms & Number of H atoms   \\
\hline
Case $1$ & $0.5$  & $0.127$ & $261,563$ & $33,147$  \\
Case $2$  & $2.5$ & $0.197$& $261,563$ & $51,398$ \\
Case $3$ & $4.5$ & $0.245$ & $261,563$ & $64,092$  \\
Case $4$ & $10.5$ & $0.387$ & $261,563$ & $101,083$  \\
Case $5$ & $17.5$ & $0.531$ & $261,563$ & $138,871$ \\
Case $6$ & $27.5$ & $0.695$ & $261,563$ & $181,830$  \\
Case $7$ & $41.5$ & $0.866$ & $261,563$ & $226,485$  \\
Case $8$ & $53.0$ & $0.958$ & $261,563$ & $250,614$ \\
\hline
\hline
\end{tabular}
\caption{Selected time steps from DMD simulations and corresponding number of atoms in MD simulations.}
\label{tab:atomnum}
\end{table}

\begin{figure}[!ht]
\centering
\includegraphics[width=1.0\textwidth]{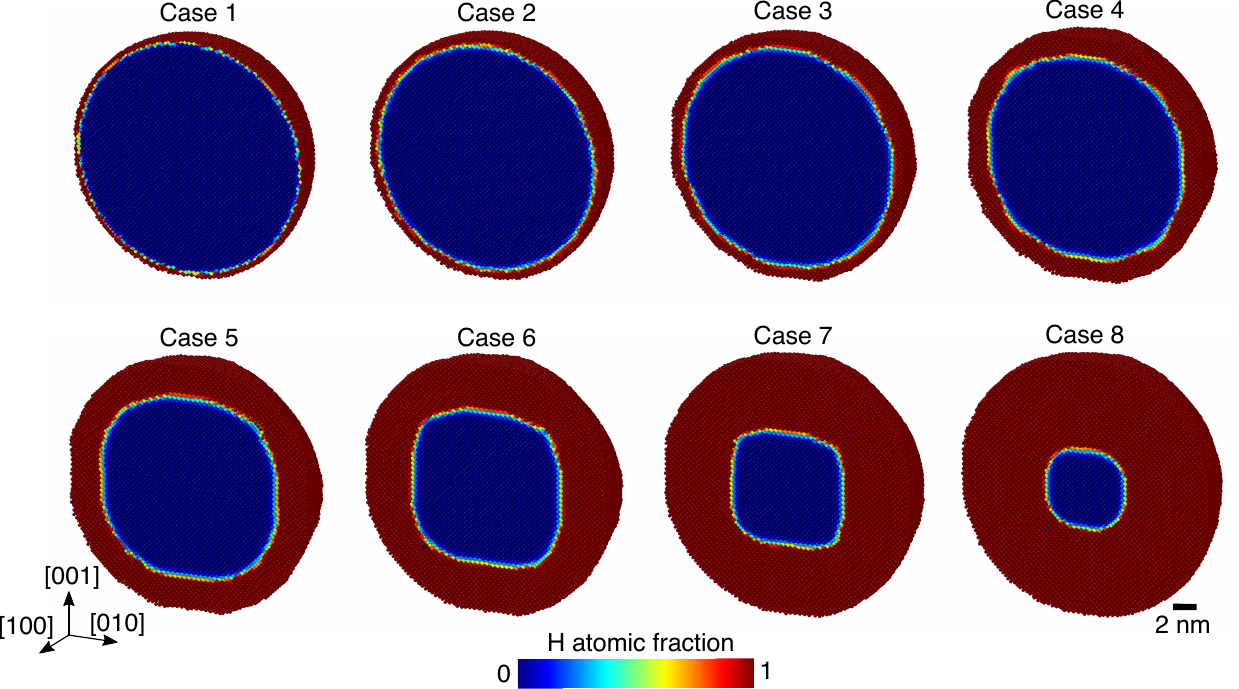}
\caption{Snapshots of H atomic fraction at the interstitial sites. Half of the nanosphere is shown.}
\label{fig:Hfraction}
\end{figure}

Fig.~\ref{fig:Hfraction} presents the evolution of H atomic fraction at interstitial sites. As may be seen, H gradually intercalated along the radial direction of the Pd nanosphere, from a spherical shell with high H concentration (close to $1$) to a core with low H concentration (close to $0$). The shell and core can be interpreted as H-concentrated $\beta$ and H-dilute $\alpha$ phase, respectively. Notably, they are separated by an abrupt boundary consisting of only a few layers of atomic sites. The thickness of such a boundary is approximately $0.5$~nm. Therefore, the phase transformation process is governed by the slow propagation of this atomistically sharp, hydride phase boundary. This mechanism is the same as experimental observations in the hydrogenation of individual Pd nanoparticles~\cite{narayan2017direct, sytwu2018visualizing}. Attendant to the $\alpha$-to-$\beta$ phase transformation, our simulation also shows that there is a lattice expansion with approximately $9.13\%$ increase in lattice constant, i.e., from $a_\text{L}=3.895~\text{\AA}$ in $\alpha$ phase to $a_\text{L}=4.274~\text{\AA}$ in $\beta$ phase. This lattice expansion subsequently results in the formation of misfit dislocations and stacking faults (SFs), which will be further discussed below.

\begin{figure}[!ht]
\centering
\includegraphics[width=1.0\textwidth]{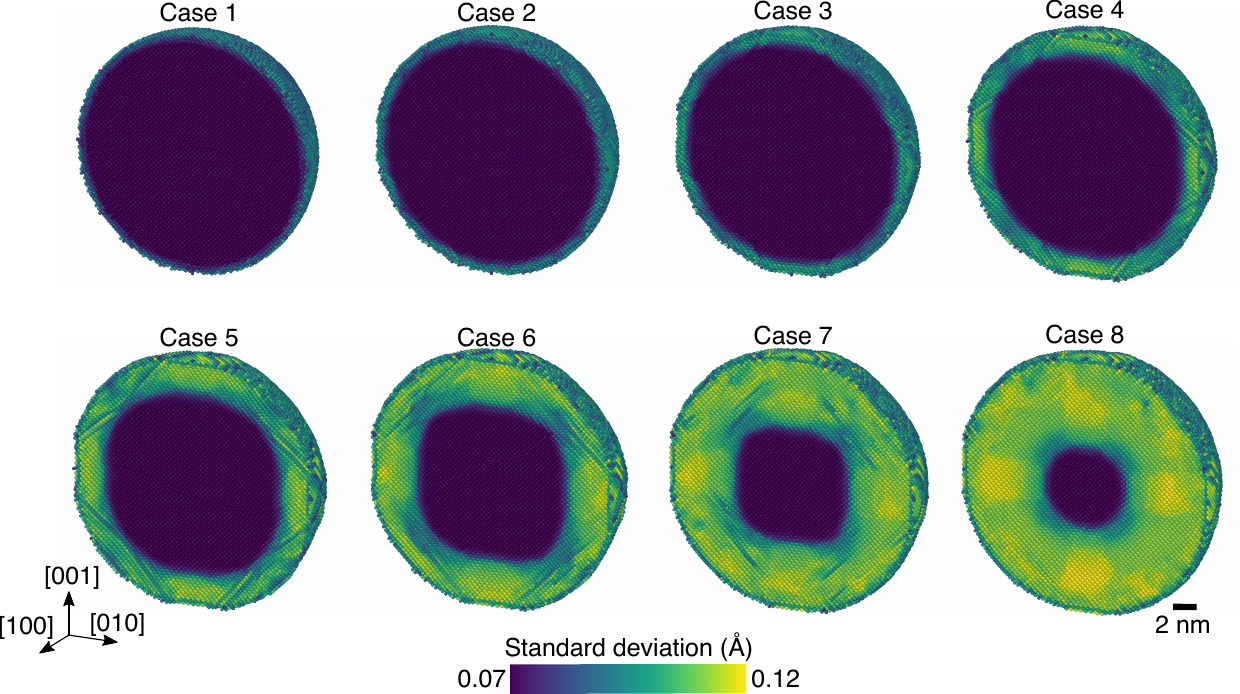}
\caption{Snapshots of SD of Pd atoms at host sites. Half of the nanosphere is shown.}
\label{fig:PdSTD}
\end{figure}
		
\begin{figure}[!ht]
\centering
\includegraphics[width=1.0\textwidth]{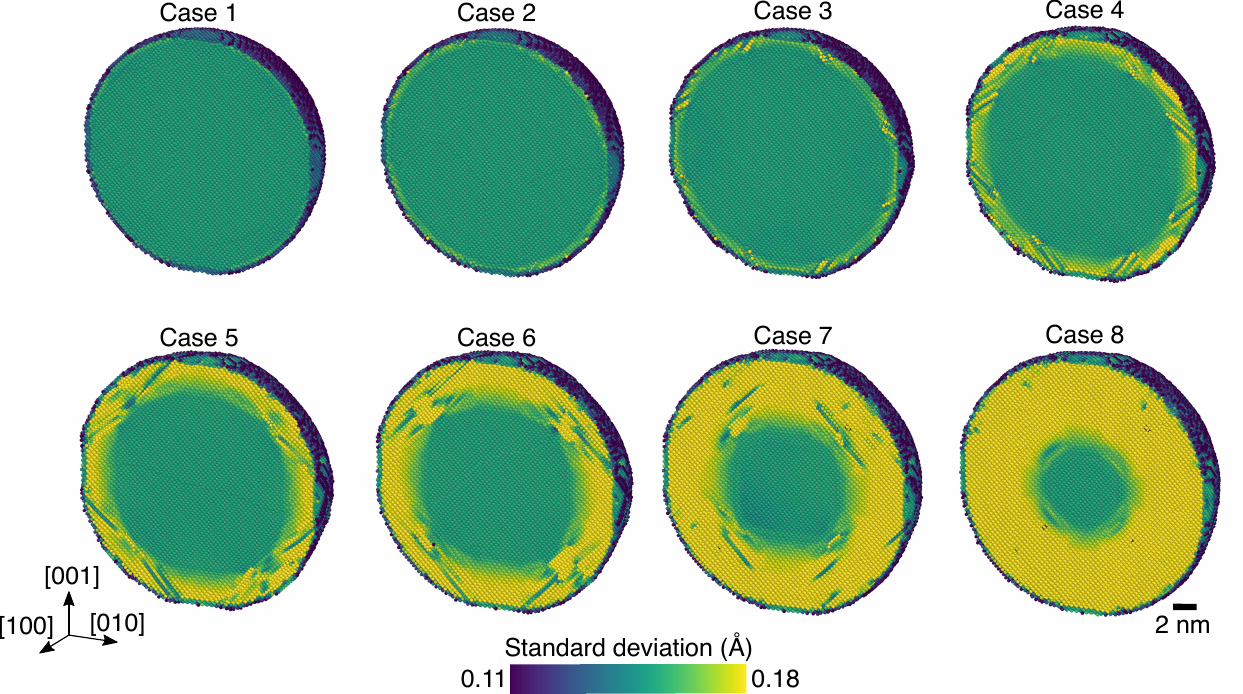}
\caption{Snapshots of SD of H atoms at interstitial sites. Half of the nanosphere is shown.}
\label{fig:HSTD}
\end{figure}

\begin{figure}[!ht]
\centering
\includegraphics[width=1.0\textwidth]{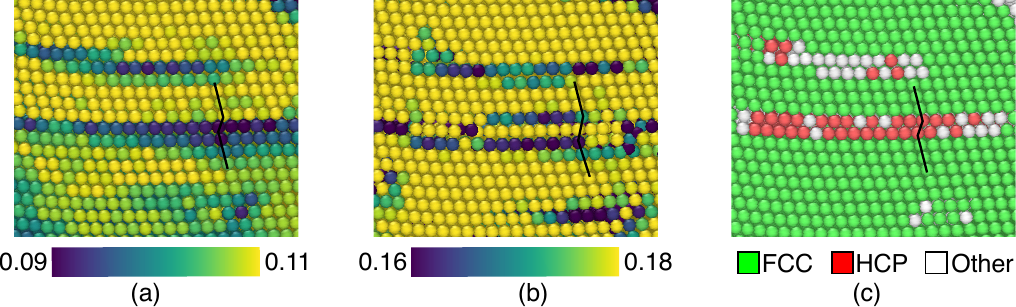}
\caption{Zoom-in views of (a) SD of Pd atoms, (b) SD of H atoms, and (c) lattice structural type in the same region near SFs. The unit in Subfigures~(a) and (b) is \AA.}
\label{fig:zoom}
\end{figure}

\begin{table}[!ht]
\centering
\begin{tabular}{ccccc}
\hline
\hline
 & \multicolumn{2}{c}{$\alpha$ phase} & \multicolumn{2}{c}{$\beta$ phase}    \\
 & Pd atom & H atom & Pd atom & H atom   \\
\hline
$\sigma_i$ (\AA) & $0.069$  & $0.153$ & $0.117$ & $0.183$  \\
%$\omega_i$ (THz) & $22.19$ & $102.84$& $13.09$ & $85.98$ \\
$\omega_i \hbar$ (meV) & $14.61$ & $67.69$& $8.62$ & $56.59$ \\
\hline
\hline
\end{tabular}
\caption{Position SD and vibrational energy of Pd and H atoms in $\alpha$ and $\beta$ phases.}
\label{tab:sig}
\end{table}
A time sequence of vibrational standard deviation (SD) of Pd and H atoms is depicted in Figs.~\ref{fig:PdSTD} and \ref{fig:HSTD}, respectively, where sites are placed based on their mean atomic positions. Similar to the hydride phase boundary, there is also a distinct and abrupt transition of SD from $\alpha$ to $\beta$ phase. To facilitate quantitative comparison, Table~\ref{tab:sig} lists the SD of atomic positions and vibrational energy in equilibrated, defect-free $\alpha$ and $\beta$ phases. As anticipated, the SD of Pd atoms is significantly lower than that of H atoms in both $\alpha$ and $\beta$ phases. In addition, the SD in the $\beta$ phase is larger than that in the $\alpha$ phase for both Pd and H atoms, owing to the stronger interactions caused by additional H atoms in the H-rich phase. Specifically, the vibrational energy of Pd atoms decreases by $41\%$ from $\alpha$ to $\beta$ phase, whereas this reduction is $16\%$ for H atoms. 

Figs.~\ref{fig:PdSTD} and \ref{fig:HSTD} also reveal that certain atoms in the $\beta$ phase exhibit lower SDs, forming a strip-shaped pattern parallel to the $\{111\}$ slip planes of FCC structures. Fig.~\ref{fig:zoom} provides zoom-in views of a representative region along with its lattice structural type. Comparing the subfigures of Fig.~\ref{fig:zoom}, we find that the regions where stacking faults (SFs) are observed coincide with the regions where atoms have lower SDs. The reduction in SD can be more than $25\%$. Thus, solute-induced SFs lead to a decrease in SD, hence increasing the vibrational energy of both Pd and H atoms. 

Moreover, Fig.~\ref{fig:HSTD} indicates that the positions of H atoms exhibit less significant deviation on the surface of the particle compared to the bulk, owing to insufficient atomic coordination. The vibrational energies of H atoms within the free surface range between $75$~meV and $115$~meV, exceeding those in the bulk. By way of comparisons, an inelastic neutron scattering study shows that with a particle size of less than $23$~nm, excess vibrational excitations of H atoms within the surface and subsurface were observed between $90$~meV and $140$~meV~\cite{stuhr1995vibrational}, which also exceeds the vibrational energy of H atoms in the bulk. This is in consistence with our DMD calculations.

\subsection{Comparisons between DMD and MD simulations}

We now proceed with the results from MD simulations. As described in Section~\ref{sec:MD}, the initial positions of Pd and H atoms for MD simulations are generated based on the means and SDs obtained by DMD simulations. We employ the statistics at the same eight time steps as shown in Figs.~\ref{fig:Hfraction}-\ref{fig:HSTD}. Consequently, a total of eight MD simulations are performed, and the numbers of Pd and H atoms in each simulation are provided in Table~\ref{tab:atomnum}.

\begin{figure}[!ht]
\centering
\includegraphics[width=1.0\textwidth]{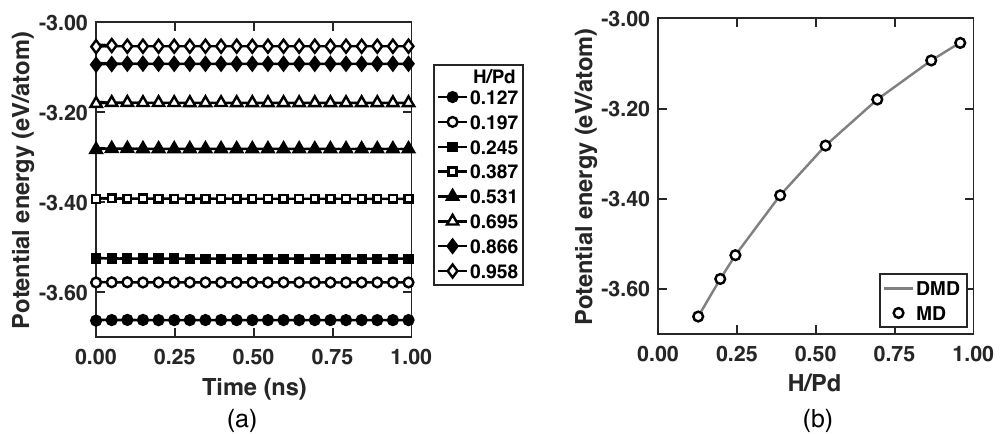}
\caption{(a) Time history of potential energy in MD simulations. (b) Comparison of potential energy between DMD and MD simulations.}
\label{fig:potential}
\end{figure}

To compare the macroscopic states obtained from the DMD and MD simulations, we focus on the potential energy per atom computed by these two methods. Fig.~\ref{fig:potential}(a) illustrates the time history of the potential energy per atom as obtained from the MD simulations. It is evident that the potential energy quickly reaches equilibrium, typically within less than 0.1 ns, for all cases examined. This indicates that the DMD simulations have successfully provided energetically stable configurations of Pd and H atoms across a wide range of H/Pd ratios. Furthermore, Fig.~\ref{fig:potential}(b) compares the time-averaged potential energy between the DMD and MD simulations. Notably, the results obtained from both methods closely align with each other, demonstrating that the averaged potential energy increases with increasing H/Pd ratio. Therefore, the atomic configurations obtained by the two methods are statistically equivalent.

\begin{figure}[!ht]
\centering
\includegraphics[width=0.8\textwidth]{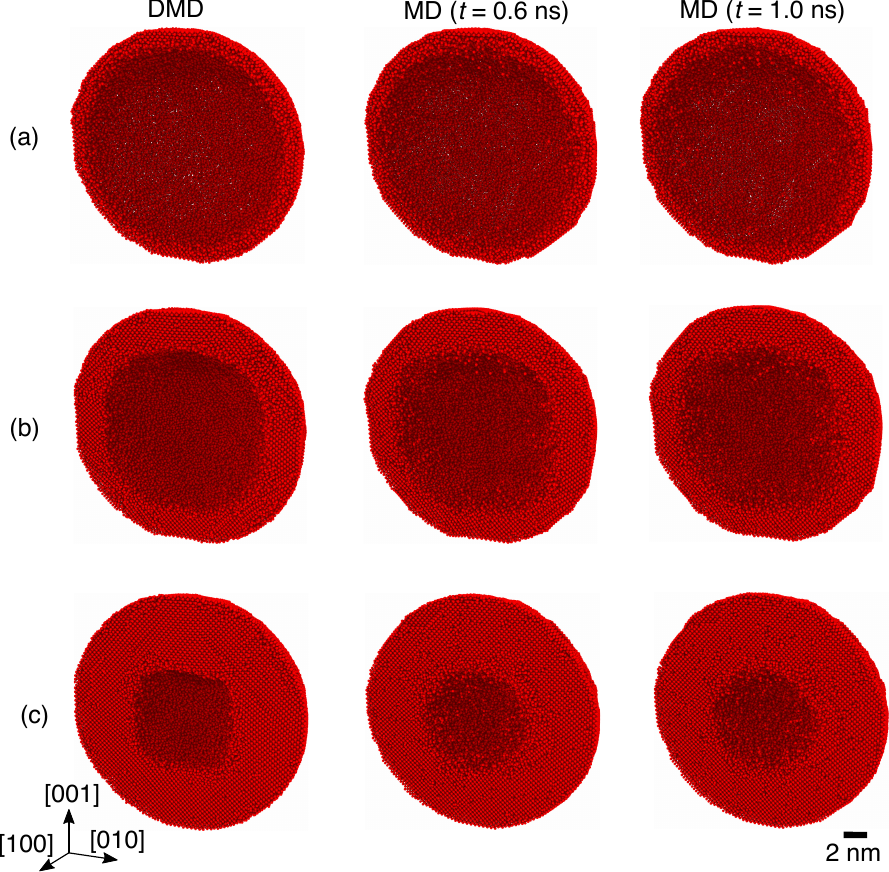}
\caption{Spatial distributions of H atoms: (a) Case 3: $\text{H/Pd}=0.245$, (b) Case 5: $\text{H/Pd}=0.531$ and (c) Case 7: $\text{H/Pd}=0.866$. In these subfigures, only half of the nanosphere is shown. The DMD results are generated by removing interstitial sites with $x_i<0.05$.}
\label{fig:phsbd}
\end{figure}

\begin{figure}[!ht]
\centering
\includegraphics[width=0.8\textwidth]{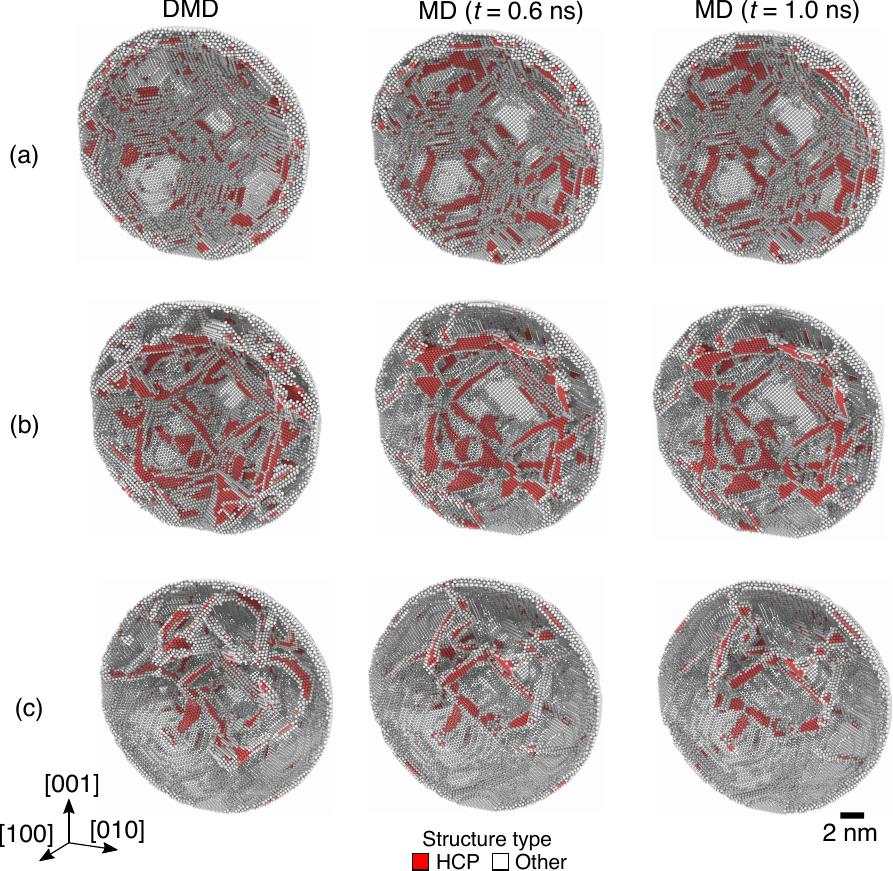}
\caption{Spatial distributions of extracted lattice defects: (a) Case 3: $\text{H/Pd}=0.245$, (b) Case 5: $\text{H/Pd}=0.531$ and (c) Case 7: $\text{H/Pd}=0.866$. In these subfigures, only half of the nanosphere is shown and the atoms with perfect FCC structure are removed for the sake of clarity.}
\label{fig:defects}
\end{figure}

%However, the MD method is not able to predict the slow motion of the phase boundary and hence the dynamics of the induced stacking faults, due to its extremely short simulation time window. By contrast, the DMD results presented in the previous sections have provided the atomic-level details involved in the entire diffusion-deformation coupled process. 

\begin{figure}[!ht]
\centering
\includegraphics[width=0.6\textwidth]{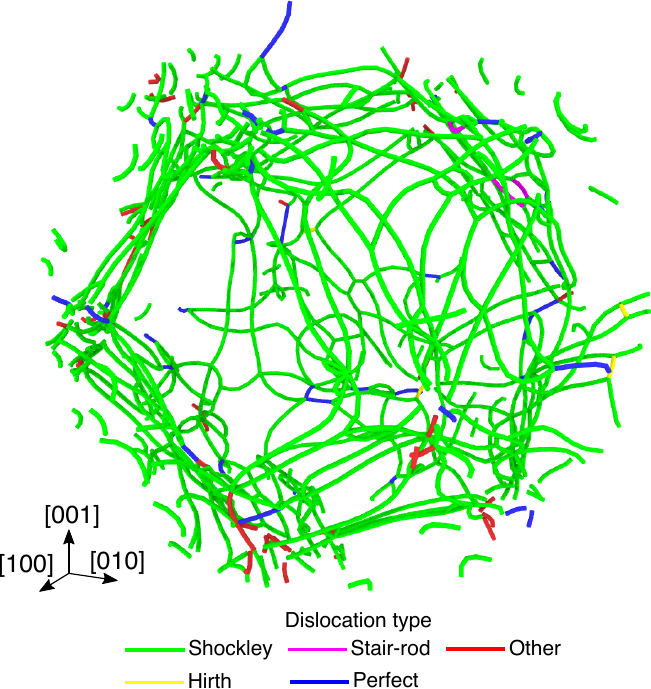}
\caption{Spatial distribution of dislocation lines at MD ($t=1.0$~ns) of Case 5: $\text{H/Pd}=0.531$.}
\label{fig:dislo}
\end{figure}

Figs.~\ref{fig:phsbd} and \ref{fig:defects} compare the spatial distributions of H atoms and SFs between DMD and MD simulations at three representative H/Pd ratios, receptively. As shown in Fig.~\ref{fig:phsbd}, MD simulations reveal that during the time window of $1.0$~ns, the equilibrium positions of most H atoms only vary slightly. As a result, the sharpness of the phase boundary is preserved, and the position coincides with that obtained from DMD simulations. This again confirms that the DMD method has provided an energetically stable configuration of atoms. Additionally, we have observed the presence of H-induced misfit dislocations and SFs, as shown in Fig.~\ref{fig:defects}. The majority of these misfit dislocations are Shockley partials, and an example of the dislocation lines is depicted in Fig.~\ref{fig:dislo}. Most of the SFs take place within the $\beta$ phase and phase boundary. They grow inwards along with the propagation of the sharp phase boundary. The formation and dynamics of SFs in the nanoparticle can be attributed to their role in alleviating the residual stress induced by the atomistically sharp boundary and the lattice mismatch between the $\alpha$ and $\beta$ phases, as previously reported by Ref.~\cite{sun2019atomistic} through an elastic core-shell model. 

In addition to the SFs revealed by the DMD simulation, the MD simulation also confirms their existence and persistence in the same locations within the short time span of $1.0$~ns. It is evident that some differences arise in the spatial distribution of lattice defects between DMD and MD simulations, as expected. It can be attributed to the fact that the MD tracks the displacive thermal vibrations of Pd and H atoms, allowing it to identify other energetically stable atomic configurations that further relax the system. On the other hand, the solution obtained from DMD represents an averaged configuration.

\begin{figure}[!ht]
\centering
\includegraphics[width=1.0\textwidth]{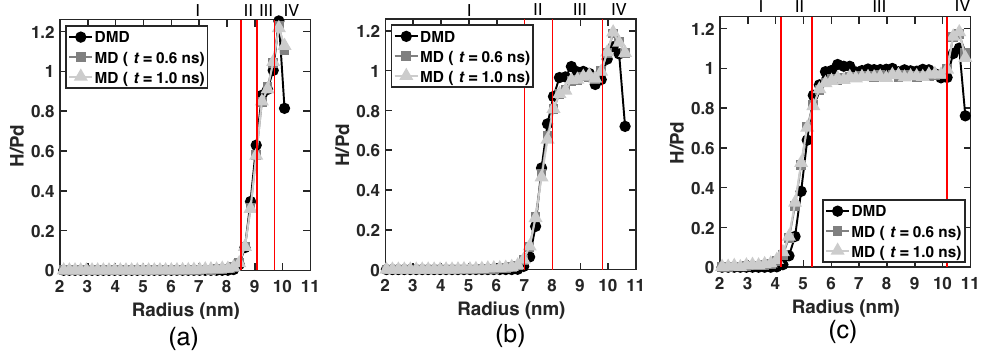}
\caption{H concentration along the radial direction of the Pd nanosphere: (a) Case 3: $\text{H/Pd}=0.245$, (b) Case 5: $\text{H/Pd}=0.531$ and (c) Case 7: $\text{H/Pd}=0.866$. In each subfigure, the four identified regions are: (I) $\alpha$ phase, (II) phase boundary, (III) $\beta$ phase and (IV) particle surface.}
\label{fig:Hconcentration}
\end{figure}

To facilitate a fair comparison and identify characteristic regions, we investigate the H distribution along the radial direction of the nanosphere. We utilize the radial H/Pd ratio, which is computed by summing the H fractions or atoms in multiple bins with a shape of spherical shell, and then dividing this sum by the number of host sites or Pd atoms within the same bin. Fig.~\ref{fig:Hconcentration} displays the radial H/Pd ratios as a function of the radius for Cases $3$, $5$ and $7$. As may be seen from the figure, it is evident that the MD simulations consistently indicate the robustness of the phase boundary within a time frame of $1$~ns. Additionally, the distributions obtained from the DMD and MD simulations overlap, further reinforcing confidence in the ability of DMD to predict energetically stable atomic configurations across a range of H/Pd ratios.

Furthermore, based on the radial H/Pd ratio, the nanosphere can be divided into four distinct regions characterized by a shape of spherical shell: (I) $\alpha$ phase ($\text{H/Pd}<0.05$), (II) phase boundary ($0.05<\text{H/Pd}<0.8$), (III) $\beta$ phase ($0.8<\text{H/Pd}<1$) and (IV) particle surface ($\text{H/Pd}>1$). The thickness of the phase boundary is around $1$~nm. This is slightly larger than the value predicted by DMD simulations, since the shape of the phase boundary is not a perfect spherical shell. Moreover, the thickness of the surface layer is around $0.7$~nm, consistent with the specified thickness for subsurface saturation in the simulation setup. The partitioning of the nanosphere into the four key parts will allow us to conduct region and feature-dependent structural analysis in what follows.

\subsection{Spacial analysis of lattice structures}

\begin{figure}[!ht]
\centering
\includegraphics[width=1.0\textwidth]{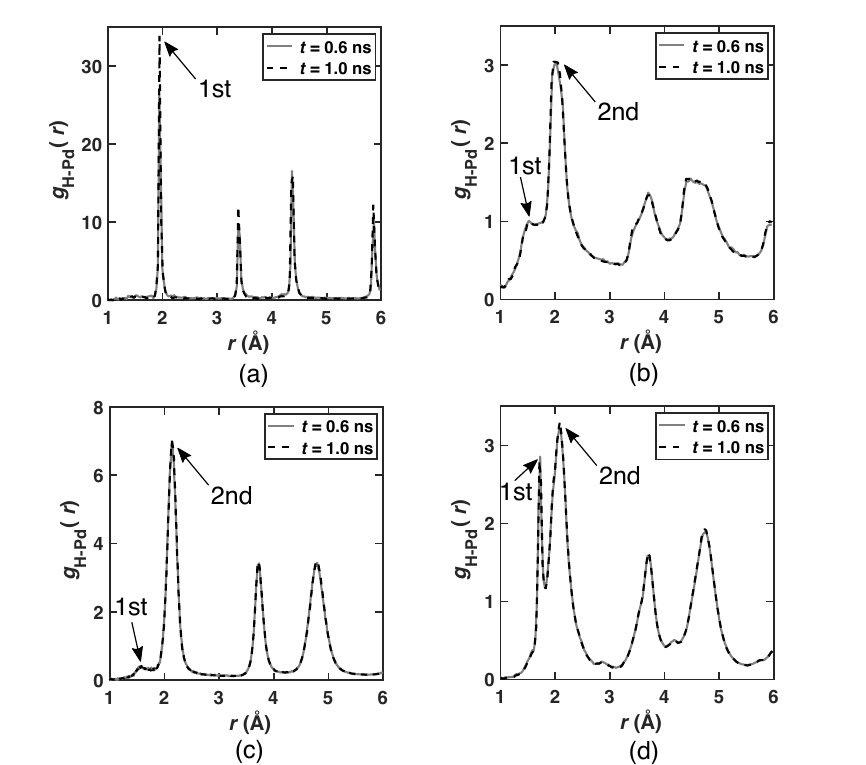}
\caption{Partial radial distribution function for H-Pd pair in the four identified regions of Case $5$: (a) $\alpha$ phase, (b) phase boundary, (c) $\beta$ phase and (d) particle surface.}
\label{fig:rdf}
\end{figure}

\begin{table}[!ht]
\centering
\begin{tabular}{ccc}
\hline
\hline
 & $1$st peak & $2$nd peak  \\
\hline
$\alpha$ phase & $1.94~\text{\AA}$ & - \\
Phase boundary & $1.64~\text{\AA}$ & $2.02~\text{\AA}$ \\
$\beta$ phase & $1.71~\text{\AA}$ & $2.14~\text{\AA}$ \\
Particle surface & $1.72~\text{\AA}$ & $2.08~\text{\AA}$ \\
\hline
\hline
\end{tabular}
\caption{Locations of the RDF peaks in the four key regions.}
\label{tab:peakloc}
\end{table}

A quantitative characterization of the identified key regions may be based on radial distribution function (RDF) analysis. For simplicity, but without loss of generality, we specifically choose the MD result of Case $5$: $\text{H/Pd}=0.531$, i.e., Fig.~\ref{fig:Hconcentration}(b), in which the volume fraction of $\alpha$ phase is approximately the same as that of $\beta$ phase. Fig.~\ref{fig:rdf} compares the RDFs of H atoms with respect to Pd neighbors in the four regions. As expected, both $\alpha$ and $\beta$ phase regions exhibit a distinct crystalline structure. In contrast, the region of phase boundary is more amorphous due to large lattice distortion. Similarly, the particle surface region also displays an amorphous nature, owing to the absence of neighboring atoms. Furthermore, using a cutoff distance of $2.4~\text{\AA}$, the RDF of the $\alpha$ phase exhibits a single peak. In contrast, there are two peaks in the RDFs of phase boundary, $\beta$ phase and particle surface. The locations of these peaks are summarized in Table~\ref{tab:peakloc}.

\begin{figure}[!ht]
\centering
\includegraphics[width=0.6\textwidth]{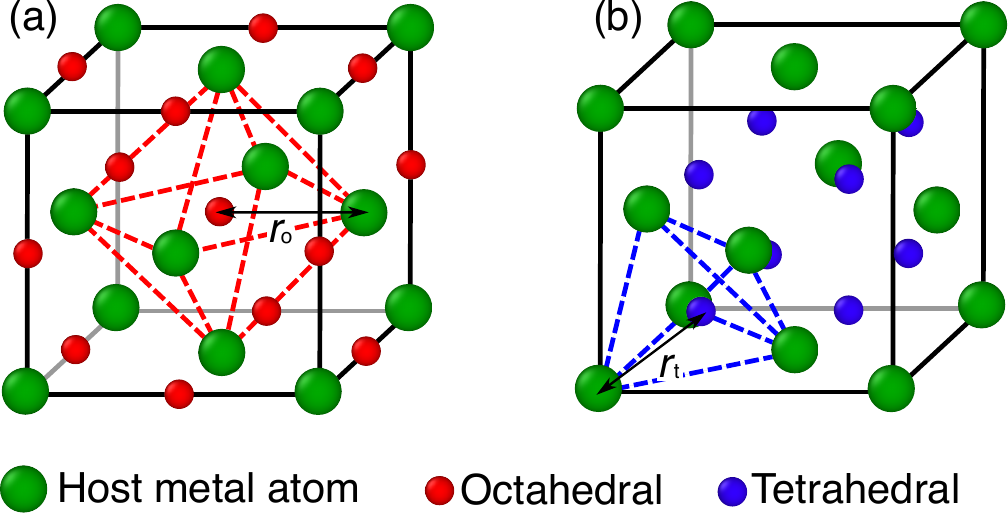}
\caption{Schematic illustration of the two types of interstitial sites in FCC lattice: (a) oct and (b) tet. $r_\text{o}$ and $r_\text{t}$ denote the distances between a metal atom and its nearest oct and tet interstitial sites, respectively. For the ideal FCC lattice without any distortion, $r_\text{o}=a_\text{L}/2$ and $r_\text{t}=\sqrt{3}a_\text{L}/4$, where $a_\text{L}$ is the lattice constant.}
\label{fig:octtet}
\end{figure} 

The difference in the number of RDF peaks between the defect-free region (i.e., $\alpha$ phase) and the other regions containing defects is of particular interest. Fig.~\ref{fig:octtet} illustrates the two types of interstitial sites present in the FCC lattice: octahedral (oct) and tetrahedral (tet) sites. In a perfect Pd lattice, the oct site is the most energetically favorable site for H occupation in both the $\alpha$ and $\beta$ phases~\cite{zhou2008embedded}. Therefore, in the $\alpha$ phase where no lattice defects are present, the first peak at $1.94~\text{\AA}$ corresponds to the distance between a Pd atom and its nearest oct sites. This explanation also accounts for the occurrence of the second peak in the $\beta$ phase and phase boundary regions. Upon closer examination, we observe that the first peak in these two defective regions may correspond to the distance between a Pd atom and its nearest tet sites. Given that most stacking faults (SFs) occur within the $\beta$ phase and phase boundary, it is plausible that the tet site is preferred over the oct site in the vicinity of SFs.

\begin{figure}[!ht]
\centering
\includegraphics[width=0.8\textwidth]{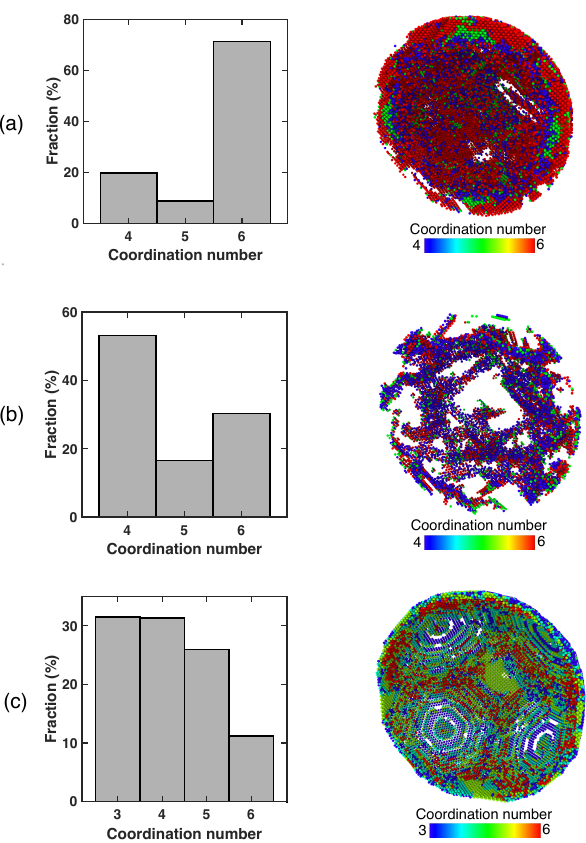}
\caption{Coordination analysis of H atoms with Pd neighbors at MD ($t=1.0$~ns) of Case 5: (a) perfect FCC lattice, (b) SFs and (c) particle surface. In each subfigure, the left figure shows the histogram of coordination numbers, and the right figure, with half of the nanosphere, shows the spatial distribution of coordination numbers. }
\label{fig:coord}
\end{figure}

To verify the preferred interstitial sites of H atoms, we perform space-dependent coordination neighboring analysis of H atoms with respect to Pd neighbors at $t=1.0$~ns in the MD simulation of Case 5. For the sake of clarity, we classify H atoms into three groups according to the types of lattice structures of their Pd neighbors: (a) perfect FCC lattice existing in $\alpha$/$\beta$ phases and phase boundary, (2) SFs existing in $\beta$ phase and phase boundary, and (3) particle surface.  Using a cut-off distance of $2.4~\text{\AA}$, the resultant coordination results are displayed in Fig.~\ref{fig:coord}. As expected, the oct site, identified by a coordination number of $6$, plays the most significant role in the perfect Pd lattice in both $\alpha$ and $\beta$ phases (see Fig.~\ref{fig:coord}(a)). In contrast, the tet site, which is 4-fold coordinated, exhibits greater prominence in the region of SFs, as shown in Fig.~\ref{fig:coord}(b). This further corroborates the preference for the tet site over the oct site in the vicinity of SFs. Furthermore, in the particle surface region, where neighbors are scarce, most of the coordination numbers have relatively small values, and the coordination numbers from $3$ to $5$ are distributed nearly uniformly.

\subsection{Stacking-fault energy}

\begin{figure}[!ht]
\centering
\includegraphics[width=1.0\textwidth]{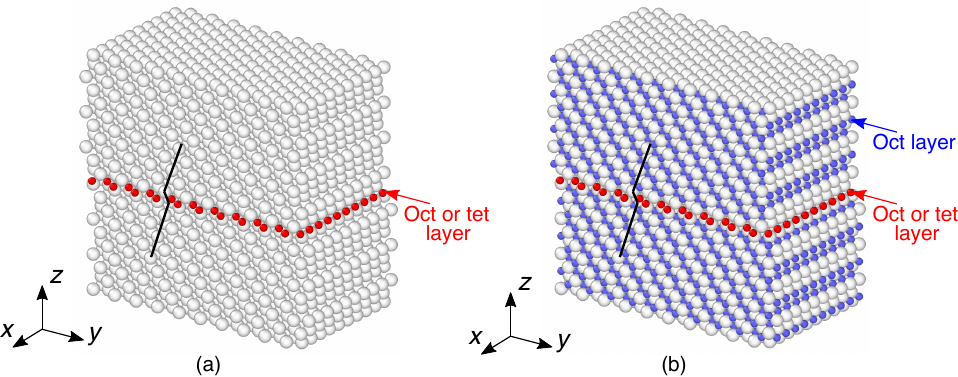}
\caption{Initial setup of the calculation of intrinsic SF energy: (a) $\alpha$ phase, and (b) $\beta$ phase. The big and small spheres denote Pd and H atoms, respectively.}
\label{fig:SF1}
\end{figure}

In order to further explore the reason why H atoms prefer to occupy the tet site within an SF and in its vicinity, we employ Molecular Statics to examine the potential energy that is induced by SFs and H atoms occupying oct/tet sites in both $\alpha$ and $\beta$ phases. Specifically we generate a simulation box consisting of both Pd and H atoms, as shown in Fig.~\ref{fig:SF1}. The $x$, $y$, and $z$ axes of the simulation box are aligned with the $[1\bar{1}0]$, $[11\bar{2}]$, and $[111]$ crystal directions, respectively. The box size is $\sqrt{6}/2a_\text{L}N_x  \times \sqrt{2}/4a_\text{L}N_y \times \sqrt{3}/3a_\text{L}N_z$ in the $x$, $y$, and $z$ directions, respectively, where $a_\text{L}$ is the lattice constant, and $N_x$, $N_y$ and $N_z$ are the numbers of Pd atomic layers in the corresponding directions. In this work, we set $N_x=10$, $N_y=20$ and $N_z=21$. The model has periodic boundaries in the $x$ and $y$ directions, while a free boundary condition is applied in the $z$ direction. 

An intrinsic SF is created by displacing the top half of the crystal as a rigid body with respect to the bottom half along the Burgers’ vector of Shockley partial dislocations. In addition to host Pd atoms, H atoms are added to the crystal to create $\alpha$ and $\beta$ hydride phases. Specifically, a $[111]$ layer of H atoms is inserted within the SF, occupying either oct or tet sites, as shown in Figs.~\ref{fig:SF1}(a) and (b). The purpose of choosing oct or tet sites is to compare the difference in SF energy between them. Then in the $\alpha$ phase, there are no additional H atoms besides those within the SF (Fig.~\ref{fig:SF1}(a)), while in $\beta$ phase H atoms occupy other oct sites in addition to those within the SF (Fig.~\ref{fig:SF1}(b)). The SF energy is calculated using the following equation
\begin{equation}
E_\text{s} = \dfrac{E_\text{f}-E_0}{A},
\label{eq:SFenergy}
\end{equation}
where $E_\text{s}$ is the SF energy, $E_\text{f}$ the potential energy after creating the SF and displacing H atoms at oct or tet sites within the SF, $E_\text{0}$ the potential energy before creating the SF and after displacing all H atoms at oct sites, and $A$ the area of the SF.

\begin{figure}[!ht]
\centering
\includegraphics[width=1.0\textwidth]{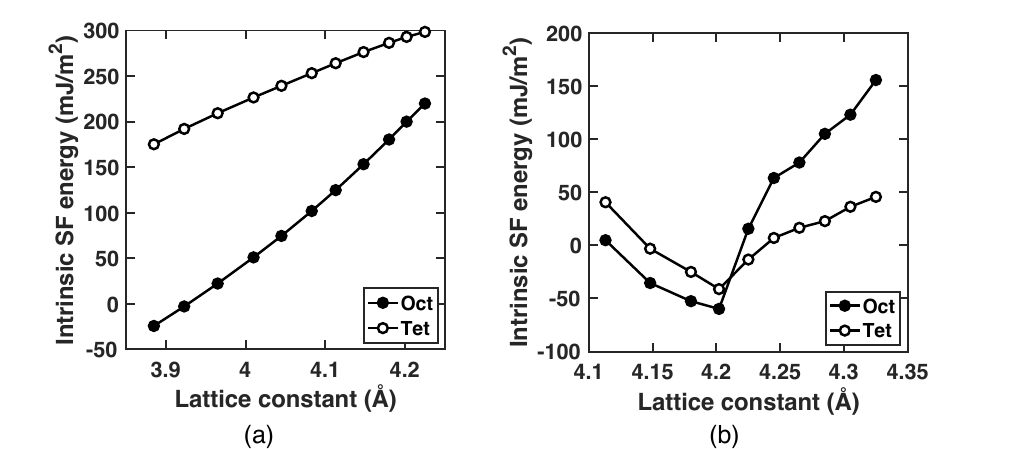}
\caption{Intrinsic SF energy as a function of initial lattice constant: (a) $\alpha$ phase, and (b) $\beta$ phase.}
\label{fig:SF2}
\end{figure}

The intrinsic SF energy has been calculated as a function of the initial lattice constant for different combinations of site type (oct or tet) within the SF and hydride phase type ($\alpha$ or $\beta$), as illustrated in Fig.~\ref{fig:SF2}. Each subfigure of Fig.~\ref{fig:SF2} corresponds to a specific value of $E_\text{0}$. Upon examination, it is evident that in the $\alpha$ phase, the SF energy of the tet site is consistently higher than that of the oct site, suggesting that the oct site exhibits greater stability within the SF. Conversely, in the H-rich $\beta$ phase, the oct site is favored when $a_\text{L}<4.2~\text{\AA}$. However, for $a_\text{L}>4.2~\text{\AA}$, the tet site demonstrates lower SF energy. The disparity in SF energy between the tet and oct sites intensifies with increasing $a_\text{L}$, indicating that the tet site becomes more energetically stable as the lattice constant becomes sufficiently large. While the exact potential energy at high H concentrations is heavily influenced by the specific local atomic environment, our findings robustly support the assertion of significantly lower SF energy induced by the tet site within H-rich environments.

\section{Concluding remarks}
\label{sec:concl}

We have conducted an analysis of hydrogen (H) absorption by a palladium (Pd) nanosphere with a diameter of $20$ nm, utilizing both Diffusive Molecular Dynamics (DMD) and classical Molecular Dynamics (MD) simulations. The key advantage of DMD, which makes it highly suitable for this study, is its capability to simulate long-term behavior of atomic systems without the time limitations inherent to MD. In contrast, MD simulations are initialized based on DMD results at various H/Pd ratios, providing more intricate insights into atomic movements and lattice relaxation within the time scale of thermal vibrations. The DMD simulations are conducted for over one second, while the time window of MD simulations spans one nanosecond.

Several significant findings from our calculations deserve attention. Firstly, the hydride phase transformation in Pd nanoparticles is primarily driven by the propagation of an atomically sharp $\alpha$/$\beta$ phase boundary. Stacking faults are observed within the $\beta$ phase and the phase boundary, effectively relieving the elastic stress arising from the lattice misfit between the $\alpha$ and $\beta$ phases. While the $\beta$ phase contributes to a decrease in vibrational energy, the presence of stacking faults locally increases the vibrational energy of both Pd and H atoms. Notably, the DMD results align closely with the MD solution in terms of time-averaged potential energy, indicating that DMD can provide energetically stable configurations of Pd and H atoms across a wide range of H/Pd ratios. Furthermore, the MD simulations confirm the stability of the hydride phase boundary and the spatial distribution of stacking faults on a timescale of nanoseconds. By analyzing the radial H concentration, we have divided the nanoparticle into four distinct regions and characterized the lattice crystalline structures within these regions through radial distribution function analysis and coordination neighboring analysis. Our observations have revealed well-defined crystalline structures in both the $\alpha$ and $\beta$ phases, while the phase boundary and free surface exhibit a more amorphous nature. We have also discovered a preference for H atoms to occupy tetrahedral interstitial sites near stacking faults due to the lower stacking fault energies offered by these sites within the H-concentrated $\beta$ phase. This underscores the remarkable bidirectional interaction between H diffusion and lattice deformation.

The present study primarily focuses on misfit dislocations and stacking faults that arise as a result of H atom diffusion. However, it is important to recognize that crystalline defects are prevalent in nanomaterials and have the potential to significantly impact the dynamic behavior of H atoms. For instance, the migration of H atoms can be enhanced along dislocations through a phenomenon known as dislocation pipe diffusion. This occurs due to a reduced activation barrier associated with lattice strain in the dislocation core and its vicinity~\cite{heuser2014direct, schiavone2016ab}. Furthermore, grain boundaries play a crucial role in mediating hydrogenation processes and can influence the storage characteristics of individual nanoparticles~\cite{alekseeva2017grain, alekseeva2021grain}. Exploring the interactions between H diffusion and these preexisting lattice defects over diffusive time scales presents itself as a promising avenue for future research.

\section*{Acknowledgments}

This work is supported by the University of Kentucky through the faculty startup fund. XS is also grateful to Profs. Michael Ortiz and Pilar Ariza, for invaluable discussions.

\bibliography{mybibfile}
\bibliographystyle{unsrt}

\end{document}